\documentclass[%
 reprint,
superscriptaddress,
 amsmath,amssymb,
 aps,
prb,
floatfix,
]{revtex4-2}

\usepackage{graphicx}
\usepackage{dcolumn}
\usepackage{bm}
\usepackage[mathlines]{lineno}
\usepackage{appendix}
\usepackage{braket}
\usepackage[dvipsnames]{xcolor}
\usepackage{amsmath,amssymb,amsfonts} 
\usepackage{makecell}
\usepackage{float}
\usepackage{bbm}
\usepackage{nicefrac}
\usepackage{wasysym}
\usepackage[normalem]{ulem}

\usepackage{hyperref}
\hypersetup{
	colorlinks=true,
	linkcolor=blue,
	filecolor=blue,
	urlcolor=blue,
	citecolor=blue,
}

\begin{document}

\title{Hierarchical Structures of Quantum Geometric Spectrum in Quasicrystals: A Renormalization-Group Study}

\author{Jundi Wang}
\thanks{These authors contributed equally to this work.}
\author{Yuxiao Chen}
\thanks{These authors contributed equally to this work.}
\affiliation{School of Physics, Peking University, Beijing 100871, China}

\author{Huaqing Huang}
\email[Contact author: ]{huaqing.huang@pku.edu.cn}
\affiliation{School of Physics, Peking University, Beijing 100871, China}%
\affiliation{Collaborative Innovation Center of Quantum Matter, Beijing 100871, China}
\affiliation{Center for High Energy Physics, Peking University, Beijing 100871, China}%

\date{\today}

\begin{abstract}
Quantum geometry, characterized by the quantum metric and Berry curvature, is a powerful framework for understanding diverse physical phenomena in quantum materials, but its behavior in non-periodic systems remains largely uncharted. Here, we uncover a universal mechanism for the divergent enhancement of the quantum metric in one-dimensional quasiperiodic systems, governed by the interplay of wavefunction criticality and spectral fractality. Using the paradigmatic Fibonacci chain, we demonstrate that the quantum metric displays a hierarchical scaling structure that mirrors the fractal organization of the energy spectrum. A real-space renormalization-group analysis yields an analytic power-law scaling, $\mathcal{G} \propto (\Delta E)^{-k}$, between the quantum metric $\mathcal{G}$ and spectral gap $\Delta E$, with the exponent $k$ dictated by the system’s self-similarity. This scaling persists in the critical Aubry-Andr\'e-Harper model but disappears in both its localized and extended phases, confirming its universality across different quasiperiodic paradigms and its unique link to criticality. Our results show that the quantum metric provides a sensitive geometric indicator of quasiperiodic criticality, and highlight quasicrystals as promising platforms for realizing unconventional giant quantum geometric effects beyond the limits of periodic crystals.
\end{abstract}

\maketitle

\textit{Introduction.}---The quantum metric tensor is a central geometric quantity in quantum mechanics that characterizes the local distance between neighboring quantum states in Hilbert space\cite{Provost1980, Resta2011}. It is the real part of the quantum geometric tensor, whose imaginary part is the well-known Berry curvature tensor. While the Berry curvature governs topological phenomena like the anomalous Hall effect \cite{RevModPhys.82.1539}, the quantum metric has recently gained prominence for its crucial role in a wide array of material properties \cite{PhysRevLett.131.240001,nwae334, yu2025quantumgeometryquantummaterials, verma2025quantumgeometryrevisitingelectronic, cheng2013quantumgeometrictensorfubinistudy}.
For example, it underlies intrinsic nonlinear conductivities in noncentrosymmetric systems, enabling novel optoelectronic responses even in topologically trivial materials \cite{gao2023quantum,wang2023quantum, PhysRevLett.131.056401, PhysRevLett.132.026301, das2023intrinsic, PhysRevX.11.011001, PhysRevX.10.041041,das2023intrinsic,ulrich2025quantumgeometricoriginintrinsic,Jiang_2025}.
Moreover, recent works show that the quantum metric contributes significantly to correlation effects, including fractional Chern insulators \cite{PhysRevB.90.165139, PhysRevLett.127.246403,PhysRevResearch.5.L032048}, exciton condensations \cite{PhysRevLett.132.236001,PhysRevB.105.L140506}, and flat-band superconductivity in moir\'e systems \cite{PhysRevB.98.220511, PhysRevLett.124.167002, PhysRevResearch.5.L012015, PhysRevB.101.060505}. Despite extensive studies in periodic and disordered systems, the behavior of the quantum metric in quasiperiodic systems remains largely unexplored.

One-dimensional (1D) quasiperiodic models provide versatile platforms for such exploration. Systems like the Fibonacci chain (FC)~\cite{PhysRevLett.50.1870} and Aubry-Andr\'e-Harper (AAH) model~\cite{aubry1980analyticity, Harper1955} exhibit exotic spectral and topological properties from deterministic aperiodicity~\cite{RevModPhys.93.045001,kraus2012topological,Kraus_2012, Tanese2014}. The FC manifests a singular continuous spectrum with multifractal eigenstates, while the generalized AAH model features tunable localization transitions and mobility edges~\cite{Ganeshan2015, Wang2020}. Crucially, topological equivalence links these models: the FC maps to a critical AAH point, both inheriting topological invariants from 2D quantum Hall ancestors~\cite{Kraus_2012}. Experimental realizations in photonic lattices~\cite{kraus2012topological}, cold atoms~\cite{Roati_2008}, and acoustic systems~\cite{Apigo_2019} further enable probes of their quantum geometry. However, the impact of their critical wavefunctions and singular continuous spectra on quantum geometry has not been systematically investigated.

\begin{figure*}
\includegraphics[width=0.8\textwidth]{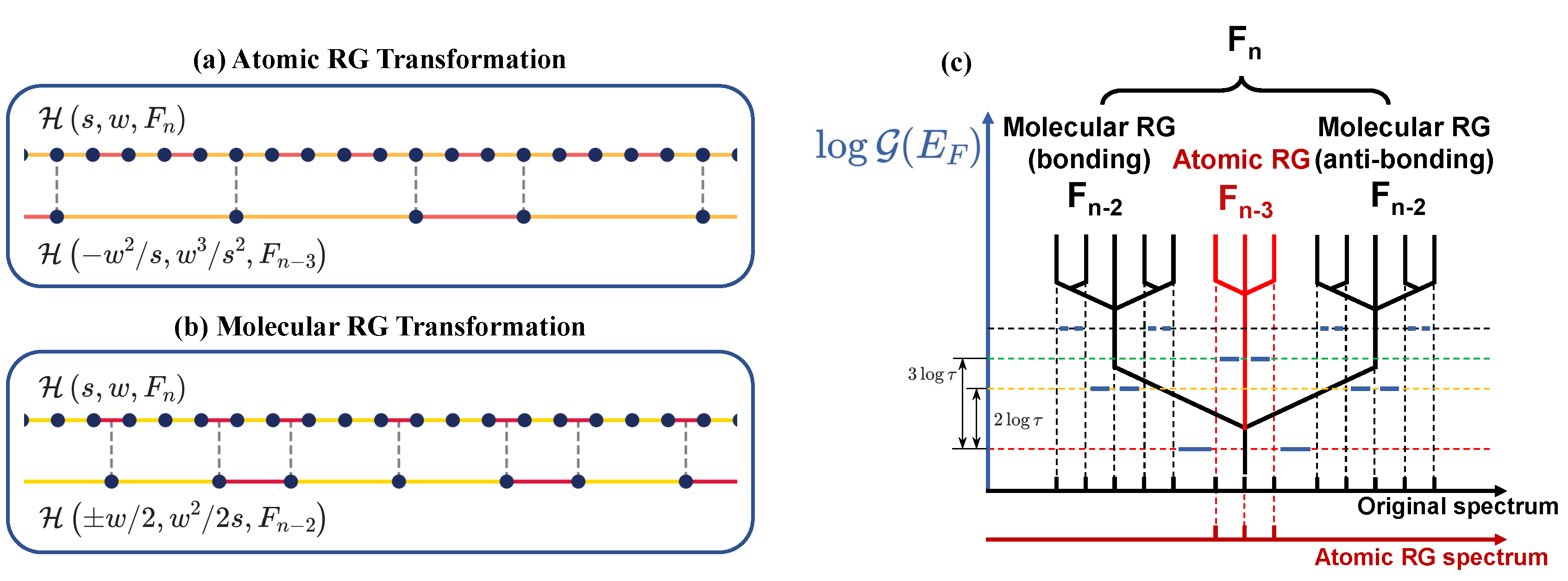}
\caption{\label{fig1} Illustration of renormalization group transformations for the FC and its fractal spectrum. (a) Atomic RG transformation (top panel), where only atomic sites (flanked by yellow weak bonds) are retained, and all other sites are decimated. These sites in the renormalized lattice are connected by effective strong and weak bonds, corresponding to pairs of atoms separated by one or two molecules in the original lattice. (b) Molecular RG transformation (bottom panel): molecules (pairs of sites connected by red strong bonds) are treated as single effective sites. Effective strong and weak bonds in the renormalized lattices are for these molecules connected by one or two weak bonds in the original lattice. (c) Schematic of the spectral splitting and the hierarchical structure of the quantum metric in the FC. Ticks on the black (red) axis denote eigenenergies for the original (renormalized) lattice. The black spectral cluster can be reproduced by the molecular RG, while the red part is reproduced by the atomic RG. The Fibonacci number $F_i$ denotes the system size. The blue vertical axis and segments represent the logarithmic quantum metric $\mathcal{G}$ versus Fermi energy $E_F$. Dashed horizontal lines indicate the hierarchical scaling of $\mathcal{G}$, corresponding to the numerical results in Fig.~\ref{fig2}(a).}
\end{figure*}

In this Letter,{we address this question by investigating the quantum metric in 1D quasiperiodic systems.} 
Prompted by the general observation that multifractal states can enhance the quantum metric \cite{PhysRevLett.133.026002}, we investigate how such geometric responses manifest in quasiperiodic critical systems. We demonstrate that wavefunction criticality in these systems leads to a universal enhancement of the quantum metric, which exhibits hierarchical scaling governed by the fractal structure of the energy spectrum. As a concrete example, we study the quantum metric spectrum of 1D off-diagonal FC. Using a real-space renormalization group (RG) framework [Fig.~\ref{fig1}(a),(b)], we derive a quantitative analytical scaling relation between the quantum metric and the size of the spectral gaps [Fig.~\ref{fig1}(c)], revealing a
{direct} connection among quantum geometry, spectral fractality, and wavefunction criticality. Our findings demonstrate that the quantum metric is significantly enhanced in quasiperiodic systems and encodes clear signatures of their underlying critical nature.

\textit{Model.}---The tight-binding Hamiltonian for the 1D off-diagonal FC can be written as \cite{PhysRevLett.50.1870, PhysRevLett.50.1873}:
\begin{equation}
\mathcal{H} = -\sum_n (t_n c_{n+1}^{\dagger}c_n+\text{h.c.}),
\label{eq:FC_H}
\end{equation}
where $c_n^{\dagger}$ ($c_n$) creates (annihilates) an electron at site $n$. The hopping amplitudes $t_n$ take two values, $t_A$ and $t_B$, arranged according to the Fibonacci sequence generated by the substitution rule $A \to AB, B \to A$ (see Supplementary Material (SM) Sec. I for details~\footnote{\label{fn}See Supplemental Material at http://link.aps.org/supplemental/xxx, for more details about the construction of the quasiperiodic models, the numerical results, and the details of perturbative renormalization group analysis, which includes Refs.~\cite{PhysRevLett.50.1870,PhysRevLett.50.1873,PhysRevLett.57.2057,Niu1990,Mace2016,aubry1980analyticity,Harper1955,janot2012quasicrystals,janssen2018aperiodic,Hofstadter1976,Ganeshan2015,Zhou2023,Marzari1997,RevModPhys.84.1419,PhysRevB.107.205133,Resta2011,yu2025quantumgeometryquantummaterials,PhysRevB.110.125124,PhysRevLett.133.026002,THOULESS197493,PhysRevLett.42.673,PhysRevA.33.1141,PhysRevLett.104.070601,PhysRevLett.103.013901,PhysRevX.14.011052,PhysRevResearch.7.023158,PhysRevLett.133.206602,PhysRevB.110.155107,superfluid2015NC,PhysRevB.98.220511,PhysRevLett.124.167002,PhysRevB.101.060505,long2025interplayquantumrealspacegeometry,Kraus_2012,kraus2012topological,bellissard1992gap,SIMON1982463,Mace_2017,PhysRevB.39.5834,ingalls1993octagonal,lin2017hyperuniformity,rhim2020quantum,PhysRevResearch.6.023256, PhysRevB.108.L140503,torma_superconductivity_2022,peotta_superfluidity_2015,PhysRevB.106.014518,PhysRevB.95.024515}.}).
\nocite{PhysRevLett.50.1870,PhysRevLett.50.1873,PhysRevLett.57.2057,Niu1990,Mace2016,aubry1980analyticity,Harper1955,janot2012quasicrystals,janssen2018aperiodic,Hofstadter1976,Ganeshan2015,Zhou2023,Marzari1997,RevModPhys.84.1419,PhysRevB.107.205133,Resta2011,yu2025quantumgeometryquantummaterials,PhysRevB.110.125124,PhysRevLett.133.026002,THOULESS197493,PhysRevLett.42.673,PhysRevA.33.1141,PhysRevLett.104.070601,PhysRevLett.103.013901,PhysRevX.14.011052,PhysRevResearch.7.023158,PhysRevLett.133.206602,PhysRevB.110.155107,superfluid2015NC,PhysRevB.98.220511,PhysRevLett.124.167002,PhysRevB.101.060505,long2025interplayquantumrealspacegeometry,Kraus_2012,kraus2012topological,bellissard1992gap,SIMON1982463,Mace_2017,PhysRevB.39.5834,ingalls1993octagonal,lin2017hyperuniformity,rhim2020quantum, PhysRevResearch.6.023256, PhysRevB.108.L140503,torma_superconductivity_2022,peotta_superfluidity_2015, PhysRevB.106.014518 PhysRevB.95.024515}
We consider the case where $t_A$ and $t_B$ correspond to weak ($w$) and strong ($s$) bonds, respectively. The FC is renowned for its singular continuous energy spectrum, which has a self-similar, Cantor-set-like structure with a hierarchy of nested gaps \cite{Niu1990}. For any non-zero quasiperiodic modulation ($w \neq s$), all its eigenstates are critical, exhibiting multifractal scaling properties that are intermediate between extended and localized states \cite{Mace2016}.

\textit{Spectral Fractality and Quantum Metric Scaling.}---To investigate the quantum geometry in systems without translational symmetry, we employ the real-space formulation of the quantum geometric tensor \cite{Resta2011, Marzari1997}:
\begin{equation}
    \mathcal{Q}_{\mu\nu}(E_F) = \text{Tr}\left[P \hat{r}_\mu (1 - P) \hat{r}_\nu P\right],
    \label{eq:Q_r}
\end{equation}
where $P = \sum_{E_\alpha < E_F} |\psi_\alpha\rangle\langle\psi_\alpha|$ is the projector onto the occupied subspace, and $\hat{r}_\mu$ denotes the position operator along the $\mu$-direction. The real part of $\mathcal{Q}_{\mu\nu}$ yields the quantum metric $\mathcal{G}_{\mu\nu}$, which quantifies the gauge-invariant spread of occupied-state wavefunctions  (See SM Sec. I.B for details~\footnotemark[\value{footnote}]). In one dimension, it reduces to a scalar $\mathcal{G}$, which can be evaluated by the interband matrix elements of the position operator \cite{Marzari1997, PhysRevLett.133.026002}:
\begin{equation}
    \mathcal{G}(E_F) =\sum_{E_\alpha < E_F < E_\beta} |\langle \psi_\alpha | \hat{x} | \psi_\beta \rangle|^2.
\end{equation}
Here, $|\psi_\alpha\rangle$ are the eigenstates with eigenvalue $E_\alpha$, and the sum runs over occupied states $\alpha$ and unoccupied states $\beta$. {The expression shows that a large quantum metric arises when occupied and unoccupied states across the Fermi level are connected by large position-operator matrix elements.}

\begin{figure*}
\includegraphics[width=0.9\linewidth]{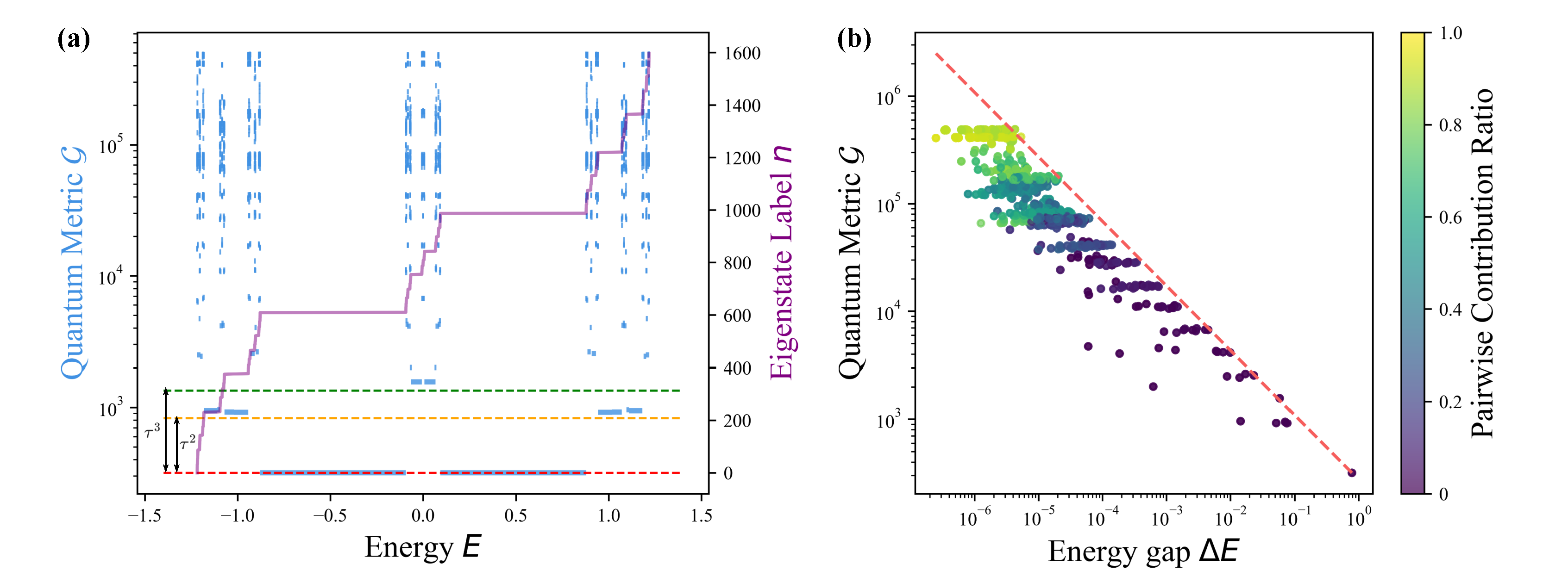}
\caption{\label{fig2} (a) Quantum metric $\mathcal{G}$ as a function of Fermi energy $E_F$ for a FC with $F_{16}=1597$ sites. The left y-axis (blue) shows $\mathcal{G}$ on a logarithmic scale; the right y-axis (purple) shows the eigenstate index $n$. Blue horizontal segments mark the value of $\mathcal{G}$ within spectral gaps, with their horizontal range and length indicating the gap’s energy range and size. {For visual clarity, gaps narrower than $\Delta E_{min}^{vis}=10^{-2}$ are plotted with this minimum display width; the corresponding $\mathcal{G}$ values are unchanged, and the actual gap sizes are used in panel (b) and in the scaling analysis.} The purple curve traces the inverted energy spectrum ($n$ vs. $E_n$). The quantum metric exhibits a hierarchical structure with scaling levels set by powers of the golden ratio $\tau = (1+\sqrt{5})/2$, as indicated by the colored dashed lines. (b) Log-Log plot revealing an inverse power-law relation between $\mathcal{G}$ and the spectral gap size $\Delta E$. The color bar indicates the contribution ratio of the eigenstate pair bordering each gap to the total quantum metric. The red dashed lines, determined by $k_{\mathrm{atomic}}$, represent the theoretical upper bound for $\mathcal{G}$. Parameters are $s=1,\ w=0.3$.}
\end{figure*}

Figure~\ref{fig2}(a) presents the quantum metric calculated for a $1597$-site ($F_{16}$) FC with off-diagonal modulation ($t_A =w= 0.3$, $t_B =s= 1$). The results reveal a fractal structure in the distribution of the quantum metric across the energy spectrum. As the Fermi energy traverses the hierarchical energy landscape, the quantum metric exhibits sharp fluctuations associated with the nested structure of spectral gaps.
The quantum metric displays a clear hierarchical scaling as the Fermi energy traverses the fractal, ternary-tree energy spectrum (as depicted in Fig.~\ref{fig1}(c)). When the Fermi energy is on a higher-level branch, the metric increases, differing by factors of $\tau^2$ or $\tau^3$ between adjacent levels---$\tau^2$ for side branches and $\tau^3$ for the middle branch---where $\tau = (1+\sqrt{5})/2$ is the golden ratio. This hierarchical organization suggests an intrinsic connection between the fractal nature of the energy spectrum and the quantum geometric properties of the wavefunctions.

To further explore this connection, we examine the relationship between the quantum metric and the spectral gap size. The results, shown by the scatter points in Fig.~\ref{fig2}(b), reveal a striking pattern: the quantum metric is significantly enhanced when the Fermi level lies within the smallest gaps of the fractal spectrum. An inverse power-law correlation $\log\mathcal{G}\sim -k\log(\Delta E)$ emerges between the gap size $\Delta E$ and the quantum metric $\mathcal{G}$, indicating a scaling behavior associated with the spectral hierarchy. The color scale indicates that for the smallest gaps, the metric's value is dominated by the contribution from the single pair of states bordering the gap. These states form bonding-antibonding state pairs with large spatial separation, leading to a large dipole matrix element $\langle\psi_{\text{occ}}|\hat{x}|\psi_{\text{unocc}}\rangle$ and thus to enhanced quantum metric (numerical verification can be found in SM Sec. III~\footnotemark[\value{footnote}]).

\textit{Renormalization-Group Theory of Quantum Metric Scaling.}---To reveal the physical origin of quantum metric enhancement in FC, we employ a perturbative real-space RG approach that leverages the system's inherent self-similarity \cite{PhysRevLett.57.2057, Mace2016}. For an off-diagonal FC with strong ($s$) and weak ($w$) bonds, we apply two distinct RG transformations that map the original FC Hamiltonian to lower-order FCs with renormalized parameters. {This procedure is analytically controlled in the strong-modulation limit $|w/s|\ll 1$, where the eigenstates 
{have dominant weight} either on isolated atomic sites or on strongly bonded molecular dimers. 
As illustrated in Fig.~\ref{fig1}(a,b), the RG transformations consist of:}
(i) An atomic RG transformation that decimates the FC from $F_n$ sites to $F_{n-3}$ sites by eliminating sites flanked by weak bonds;
(ii) Two molecular RG transformations (bonding and antibonding) that reduce the chain to $F_{n-2}$ sites by replacing strongly bonded dimers with effective sites.

These transformations yield well-defined recursion relations for the renormalized hopping amplitudes:
\begin{eqnarray}
    &\mathcal{H}(s, w,F_n) \xrightarrow{\makebox[45pt]{$\mathrm{atomic}$}} &\ \mathcal{H}(-w^2/s, w^3/s^2,F_{n-3}) \; \label{atomic}\\
    &\mathcal{H}(s, w,F_n) \xrightarrow{\makebox[45pt]{$\mathrm{molecular}$}} &\ \mathcal{H}(\pm w/2, w^2/2s,F_{n-2}) \; \label{molecular}
\end{eqnarray}
where $\mathcal{H}(s, w,F_n)$ is the Hamiltonian on an FC lattice with $F_n$ sites, $s$ and $w$ represent strong and weak hopping amplitudes, respectively. The sign in the molecular case corresponds to whether the initial molecular state is bonding ($+w/2$) or antibonding ($-w/2$). Because the ratio of two consecutive Fibonacci numbers $\tau_n=F_{n-1}/F_{n-2}$ tends to the golden mean in the large $n$ limit: $\lim_{n \rightarrow \infty}\tau_n=\tau=(1+\sqrt{5})/2$, the system size $N$ of the renormalized system reduces by a factor of $F_{n-3}/F_n=1/(\tau_{n-1}\tau_n\tau_{n+1})\approx1/\tau^3$ and $F_{n-2}/F_n\approx1/\tau^2$ for the atomic and molecular RG transformations, respectively \cite{RevModPhys.93.045001}.

These RG operations generate the fractal structure of the FC energy spectrum. As shown in Fig.~\ref{fig1}(c), the atomic RG preserves the central spectral region (red), while the molecular RG reproduces peripheral regions (black). Iterative application generates hierarchical gap structures. Owing to self-similarity, repeated application of these operations reveals that each spectral cluster contains subclusters separated by higher-order gaps, resulting in the characteristic fractal spectrum of the FC.

A key observation is that each RG step preferentially decimates lattice sites where the corresponding eigenstate has negligible amplitude. In principle, we can reproduce the dominant features of the eigenstates by introducing a wavefunction reconstruction method (see Sec. IV in SM for details~\footnotemark[\value{footnote}]). Since the RG transformations preserve wavefunction structure, they necessarily preserve quantum metric properties as well. This preservation enables us to establish a hierarchical scaling relation for the quantum metric:
\begin{equation}
\begin{aligned}
    \mathcal{G}(E_F;\mathcal{H}(s,w, F_n), a) &=\mathcal{G}(E_F; \mathcal{H}(\mathrm{R}s,\mathrm{R}w,F_{n-m}),\tau^{m} a)\\
    &=\tau^{2m}\mathcal{G}(E_F; \mathcal{H}(\mathrm{R}s,\mathrm{R}w,F_{n-m}),a)\\
    &=\tau^m\mathcal{G}(E_F; \mathcal{H}(\mathrm{R}s,\mathrm{R}w,F_{n}),a)\\
    &=\tau^{m}\mathcal{G}(\mathrm{R}^{-1}(E_F); \mathcal{H}(s,w,F_{n}),a),
\end{aligned}
\label{qmrg}
\end{equation}
where $\mathcal{G}(E_F;\mathcal{H}(s,w, F_n), a)$ is the quantum metric evaluated at Fermi energy $E_F$ from the Hamiltonian $\mathcal{H}(s,w, F_n)$ on a FC of $F_n$ sites with bond length $a$. R$s$ and R$w$ denote the renormalized hopping amplitudes under atomic or molecular RG transformations. The second line applies the quantum metric's length squared scaling under the bond-length rescaling $a\rightarrow \tau^m a$. The third line of Eq.~\eqref{qmrg} is justified by the relation derived from the linear scaling of the quantum metric with system size (proven in SM Sec. V~\footnotemark[\value{footnote}]): $\mathcal{G}(E_F,F_n)=\tau^m\mathcal{G}(E_F,F_{n-m}),\label{recurrence}$ where $m=2$ or 3 for molecular and atomic RG steps, respectively. Finally, we perform an energy rescaling for the Hamiltonian in the fourth line and transform the Fermi energy correspondingly $E_F\rightarrow\mathrm{R}^{-1}(E_F)$ to preserve the filling fraction. This is accomplished by multiplying the Hamiltonian and the Fermi energy by a suitable factor ($s^2/w^2$ for the atomic RG and $2s/w$ for the molecular RG), and performing a local gauge transformation to eliminate any sign change in the hopping terms (as shown in SM Sec.~V.B~\footnotemark[\value{footnote}]). During this ``inverse" RG procedure, the relative spectral hierarchical structure is maintained and the quantum metric remains unaffected since $\mathcal{G}$ is invariant under either energy rescaling or gauge transformation. Thus, Eq.~\eqref{qmrg} reveals a direct scaling relation between quantum metrics at different Fermi levels in the same FC, highlighting the deep correspondence between the hierarchical structures of the quantum metric and the energy spectrum.

\begin{figure*}[htb]
    \includegraphics[width=0.8\linewidth]{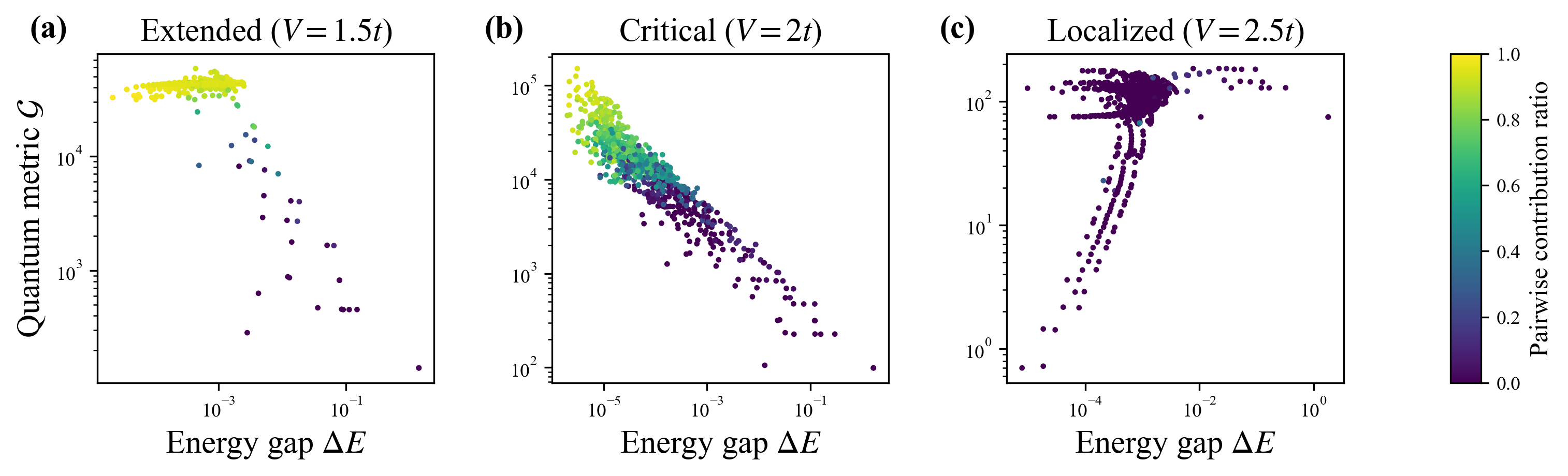}
    \caption{\label{fig11} Quantum metric-gap relationship across different regimes of the AAH model. The quantum metric versus energy gap is plotted for three distinct parameter regimes: (a) extended phase ($V=1.5t$), (b) critical point ($V=2t$), and (c) localized phase ($V=2.5t$). Color coding indicates the contribution ratio of the eigenstate pair bordering each gap to the total quantum metric.
    }
    \end{figure*}

This framework allows us to analytically derive the relationship between the quantum metric and the spectral gap size. For atomic and molecular RG transformations, the renormalized spectral gap and quantum metric transforms as $\Delta E'=\frac{w^2}{s^2}\Delta E$, $\mathcal{G}'=\tau^3 \mathcal{G}$ and $\Delta E'=\frac{w}{2s} \Delta E$, $\mathcal{G}'=\tau^2 \mathcal{G}$, respectively. Taking logarithmic derivatives $k=\frac{d\log \mathcal{G}}{d\log\Delta E}$ yields negative slopes:
\begin{equation}
    k_{\mathrm{atomic}}= \frac{3\log \tau}{2\log \frac{w}{s}},\label{qmvse}\quad
    k_{\mathrm{molecular}}= \frac{2\log \tau}{\log \frac{w}{2s}}.
\end{equation}
These results establish a power-law scaling relation $\mathcal{G} \propto (\Delta E)^k$, providing a rigorous theoretical explanation for the observed enhancement of the quantum metric at small spectral gaps in the FC. As shown in Fig.~\ref{fig2}(b), each scatter point can be connected to the minimal $\mathcal{G}$ at the largest gap $\Delta E$ (the rightmost point) via a sequence of RG steps, where each step follows either the atomic or molecular scheme. Particularly, the red dashed line with a slope of $\min(k_{\mathrm{atomic}},k_{\mathrm{molecular}})$ marks the theoretical upper bound for $\mathcal{G}$ at spectral gap $\Delta E$, which aligns well with the numerical results [see Fig.~\ref{fig2}(b)]. This scaling behavior emerges directly from the self-similar structure of the system captured by the RG framework, linking quantum geometry to the hierarchical organization of the energy spectrum. Although the analytical derivation relies on the perturbative limit $|w/s| \ll 1$, numerical results indicate that the inverse power-law scaling between the quantum metric and the spectral gap persists even when the system moves away from this limit (e.g., $w/s \to 1$), reflecting the robustness of quasiperiodic critical multifractality (see Sec. VI.B in SM for details~\footnotemark[\value{footnote}]).

\textit{Universality of Quantum Metric Scaling in Critical Quasiperiodic Systems.}---To investigate whether the quantum metric scaling observed in the FC is a universal feature of critical systems, we examine the AAH model, another paradigmatic quasiperiodic system but with fundamentally different origins. Unlike the FC, which is generated by substitution rules, the AAH model introduces quasiperiodicity through an incommensurate cosine potential \cite{aubry1980analyticity, Harper1955}:
\begin{equation}
    H = \sum_n (t c_{n+1}^{\dagger}c_n+\text{h.c.})+V\cos(2\pi \omega n+\phi)c_n^{\dagger}c_n, \label{eq:AAH}
\end{equation}
where $t$ is the nearest-neighbor hopping amplitude, $V$ controls the potential strength, $\phi$ is a phase factor, and $\omega$ is an irrational number (typically the golden ratio) ensuring incommensurability. A remarkable feature of the AAH model is its self-duality at $V=2t$, which separates extended states ($V<2t$) from localized states ($V>2t$).

Figure~\ref{fig11} compares the quantum metric across these three regimes. In the extended ($V=1.5t$) and localized ($V=2.5t$) phases, the quantum metric shows no significant correlation to gap size. In stark contrast, precisely at the critical point ($V=2t$), we observe a pronounced inverse relationship between the quantum metric and gap size---remarkably similar to the pattern observed in the FC. This striking correspondence occurs despite the different origins of quasiperiodicity in these models, suggesting that quantum metric enhancement near small spectral gaps is a universal signature of criticality rather than a model-specific feature.

The absence of this scaling relationship in both the extended and localized phases confirms its association with criticality. In the extended phase, states remain extended across all energy scales without a hierarchical structure, while in the localized phase, the exponential decay of wavefunctions suppresses long-range correlations necessary for quantum metric enhancement. Only at criticality does the perfect balance between localization and delocalization generate the hierarchical structure of states with power-law spatial correlations that yield the observed quantum metric scaling.

From an RG perspective, this universality stems from the fact that both the FC and the critical AAH model correspond to fixed points characterized by self-similarity and scale invariance. At criticality, RG transformations preserve the structural form of the Hamiltonian while rescaling energy and length scales multiplicatively. This scale invariance ensures that energy gaps and spatial correlations evolve in a coordinated fashion: as gaps become smaller through successive RG iterations, the associated wavefunctions become increasingly extended with stronger spatial correlations, leading to enhanced dipole matrix elements and consequently larger quantum metric contributions.

\textit{Conclusion.}---{We have highlighted a direct connection 
between quantum geometry, spectral fractality, and criticality in 1D quasiperiodic systems.} We showed that the quantum metric is significantly enhanced in these systems due to the long-range correlations of critical wavefunctions. In the FC, this enhancement follows a hierarchical scaling law, which we derived analytically using a real-space RG framework. This scaling behavior, characterized by a power-law relationship between the quantum metric and the spectral gap size, is shown to be a universal feature of criticality, as further evidenced by its appearance in the critical AAH model. To establish a connection to a concrete physical observable, we compute in Sec. VII.C of the SM~\footnotemark[\value{footnote}] the superfluid stiffness of the superconductivity in the FC, which exhibits a hierarchical oscillatory structure that closely follows that of the quantum metric.
Our work, therefore, demonstrates the quantum metric as {an additional and complementary geometric indicator of critical phenomena alongside more established diagnostics such as multifractality and wave-function-dynamics measurements}, and suggests that quasiperiodic systems are promising platforms for engineering novel quantum geometric effects beyond the paradigm of periodic crystals. Additional details, including further numerical analysis and references supporting the methods used, are provided in SM~\footnotemark[\value{footnote}].

\begin{acknowledgments}
We thank Prof. Anuradha Jagannathan for valuable comments and suggestions. This work is supported by the National Key R\&D Program of China (Grant No. 2021YFA1401600), the National Natural Science Foundation of China (Grant No. 12474056), and the 2022 basic discipline top-notch students training program 2.0 research project (Grant No. 20222005). The work was carried out at the National Supercomputer Center in Tianjin, and the calculations were performed on Tianhe new generation supercomputer. The high-performance computing platform of Peking University supported the computational resources.

\textit{Note added}: After completion of this work, we became aware of two related preprints posted on arXiv. Marsal \textit{et al.} \cite{marsal2025quantummetriclocalizationquasicrystal} investigated quantum-metric and localization properties in quasiperiodic chains using a mixed position–phason formulation. Sun  \textit{et al.} \cite{sun2025geometricsuperfluidweightquasicrystals} studied the superfluid weight in quasiperiodic flat-band models and related it to a flux-space quantum metric. These works address complementary aspects of quasiperiodic quantum geometry and were carried out independently of the present study.

\end{acknowledgments}

%

\end{document}